\documentclass[a4paper,11pt]{article}
\usepackage{jcappub} 

\usepackage{lineno}
\usepackage{graphicx}
\usepackage{caption}
\usepackage{subcaption}
\usepackage{hyperref}
\usepackage{cancel}

\arxivnumber{2502.08876} 

\title{\boldmath DESI Dark Secrets}

\author[1,2,\dagger]{Matilde L. Abreu\note[$\dagger$]{Present address:
                Department of Physics, Harvard University,
                17 Oxford Street, Cambridge, MA 02138, USA}}
\author[1,3]{Michael S. Turner}

\affiliation[1]{Department of Physics and Astronomy,\\
                University of California, Los Angeles,\\
                Los Angeles, CA 90095-1547, USA}
\affiliation[2]{Institute of Astronomy,\\
                University of Cambridge,\\
                Cambridge, CB3 0HA, UK}
\affiliation[3]{Kavli Institute for Cosmological Physics,\\
                University of Chicago,\\
                Chicago, IL 60637-1433, USA}

\emailAdd{tildabreu@g.ucla.edu, mturner@uchicago.edu}

\abstract{
We critically examine the results of the Dark Energy Spectroscopic Instrument (DESI) and the evidence that dark energy may not be quantum vacuum energy.  We find that the best-fit $w_0w_a$ models for dark energy, which underpin the claim, have unusual behavior.  They achieve a maximum energy density around $z\simeq 0.5 $ and rapidly decrease before and after. This redshift, where $w =-1$, is also the pivot point for the DESI results.  
We show that all this could be explained by the limited ability of the $w_0w_a$ parameterization to model dark energy
since $w_a \not= 0$ and $w=-1$ can only be achieved at a minimum or maximum of the dark energy.  Because $w_0w_a$ is a one-parameter characterization of scalar-field models it cannot represent them to the precision needed for the DESI results, and we explore scalar-field models characterized by one dimensionless parameter $\beta$, which for $\beta \rightarrow 0$, reduce to $\Lambda$CDM.  None of these models fit the DESI data significantly better than $\Lambda$CDM or as well as the best DESI $w_0w_a$ models, though they favor $\Omega_M \simeq 0.3$ rather than the higher values preferred by $w_0w_a$ models.   We also examine the CMB and SN data that strengthen the DESI case for evolving dark energy. The combination of DESI, CMB, and SN data favors a 95\% credible interval $\beta = 0.33 - 0.96$, providing weak evidence for a scalar-field explanation for dark energy.
While the DESI data prefer $w_0w_a$ to a scalar field, the SN data prefer a scalar field to $w_0w_a$.
Finally, the unusual behavior of the favored DESI $w_0w_a$ models could arise due to the matter density not varying as expected or an unaccounted-for component of energy density in the Universe.  
The evidence for evolving dark energy is intriguing but not conclusive, and at this point, the DESI results raise more questions than they answer.}

\begin{document}
\maketitle
\flushbottom

\section{Introduction}

The percent-level measurements by DESI of distances out to redshifts beyond two, combined with CMB and SN data, provide evidence at $3\sigma$ to $4\sigma$ that dark energy varies and is not quantum vacuum energy ($\Lambda$).  If true, this would be an extraordinary development in the more than 25-year quest to understand cosmic acceleration.  
However, the evidence, based upon $w_0w_a$ models for dark energy, raises many questions. For example, the best-fit models achieve a maximum energy density around $z\simeq 0.5 $ and rapidly decrease before and after.   Such behavior does not correspond to any simple physical model for dark energy.

Given the importance of the DESI results, our paper aims to better understand and to scrutinize the evidence for time-varying dark energy.  In particular, we analyze the $w_0w_a$ models that underpin the claims, explore physically motivated models for varying dark energy involving a rolling scalar-field, critically examine the DESI results and their relationship with the SN data, and discuss alternative explanations.  

The DESI results have rightly attracted much attention and scrutiny \cite{2404.14341,2404.05722,2405.03933,2405.17396,2405.18747,2408.07175,2502.06929,2505.18937,2506.13047,2506.21542,2507.03090,2408.17318}. Most studies that consider scalar-field dark energy models find them comparable to or slightly better than $\Lambda$CDM, though generally not as good as the best $w_0w_a$ models \cite{2404.14341,2405.17396,2405.18747,2502.06929,2505.18937,2506.13047,2506.21542,2507.03090}.  Some studies find no significant evidence for evolving dark energy \cite{2405.03933,2408.17318}. We discuss the relationship between our results and other contemporary work in the concluding remarks.

\subsection{Describing dark energy}
Dark energy can be characterized by an equation-of-state (EOS) parameter $w$ \cite{TurnerWhite}, which may or may not evolve with time, and the evolution of its energy density is given by,
\begin{equation} \label{eq1}
d \ln \rho_{DE} = -3(1+w) d \ln a ,
\end{equation}
where $a$ is the cosmic scale factor.  In the case that $w$ is constant, $\rho_{DE} \propto a^{-3(1+w)}$.  

For the current cosmological paradigm, $\Lambda$CDM, the dark energy is quantum vacuum energy with an unvarying EOS parameter $w=-1$, so that $\rho_{DE} = $ const.  The data \cite{PlanckLegacy,Pantheon,DES5}, including the recent DESI results \cite{DESI,DR2}, are consistent with $\Lambda$.   However, there is no compelling theoretical reason to prefer this hypothesis, and quantum vacuum energy can be falsified by showing that $w \not= -1$, that $w$ evolves with time, or both.  

A standard parameterization to test for varying dark energy is ``$w_0w_a,$",
\begin{eqnarray}
 w  =  w_0 + w_a(1-a) = w_0 +w_a z/(1+z)  =  -1 +\alpha -w_a a ,
 \end{eqnarray}
where $\alpha \equiv 1+w_0+w_a$, $a = 1$ today, and the cosmic scale factor and cosmological redshift are related by: $a(t) = 1/(1+z)$.   
It follows from Eq.~\ref{eq1} that the energy density of dark energy is a power-law times an exponential in the scale factor $a$, 
\begin{equation} \label{eq3}
\rho_{DE} \propto a^{-3\alpha} \exp [3w_a (a-1)].
\end{equation}
where $\alpha$ sets the power-law index ($-3\alpha$) and $w_a$ sets the exponential behaviour ($3w_a$).  These models, which are fully characterized by $\alpha$ and $w_a$, are discussed in more detail in \S\ref{w0wa_revisited}.

\subsection{Evidence for evolving dark energy}

The DESI Collaboration has now published two sets of BAO distances: DR1 \cite{DESI}, the first year results, and DR2 \cite{DR2}, the three-year results.  The three-year results (DR2), comprise more than 19 million redshifts and 13 distances with 0.5\% to 2.5\% precision \cite{DR2}, and significantly extend their first-year results, 12 BAO distances from a sample of 6.4 million redshifts with 1.3\% to 3\% precision \cite{DESI}. 

DR1 and DR2 are consistent with one another and alone provide $2\sigma$ evidence for evolving dark energy.\footnote{The DES 5-year data \cite{DES5} also show indications of evolving dark energy.}  Combining the DESI results with CMB and SN data \cite{PlanckLegacy,DES5,Pantheon}, the significance rises to around $3\sigma$, for a model with parameters $\Omega_M \simeq 0.3, w_0 \simeq -0.7$ and $w_a \simeq -1$, cf. Table V and Fig.~11 in Ref.~\cite{DR2}.  Throughout, we will use this model as a baseline and refer to it as the DESI+ $w_0w_a$ model.

\begin{figure}[tbp]
\center\includegraphics[width = 0.75\textwidth]{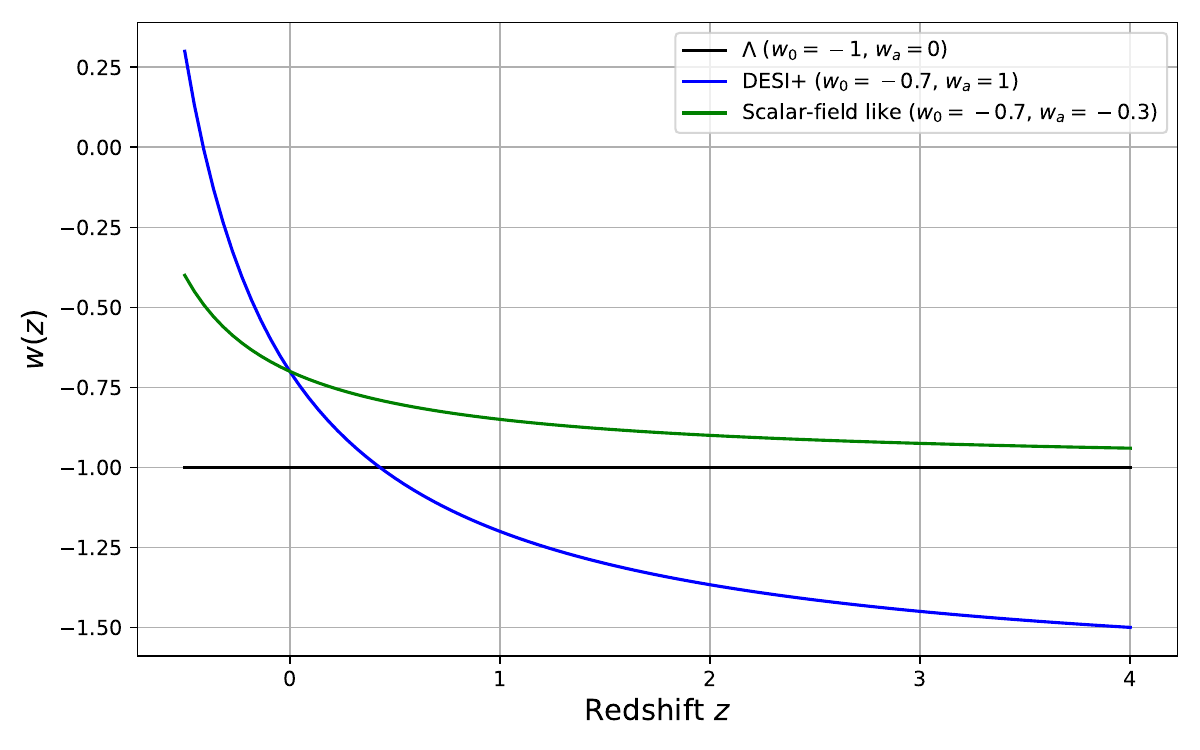}
\caption{Dark energy EOS vs. redshift for three models:   $\Lambda$ ($w_0 = -1$, $w_a =0$); behavior expected for a scalar field ($w_0 = -0.7$, $w_a = -0.3$); and the DESI+ ($w_0 = -0.7$, $w_a = -1.0$). The DESI+ model crosses the phantom divide ($w=-1$) around the pivot point ($z \simeq 0.5$) where $\rho_{DE}$ has a maximum.}

\label{EOSw0wa}     
\end{figure}

\begin{figure}[tbp]
\center\includegraphics[width = 0.75\textwidth]{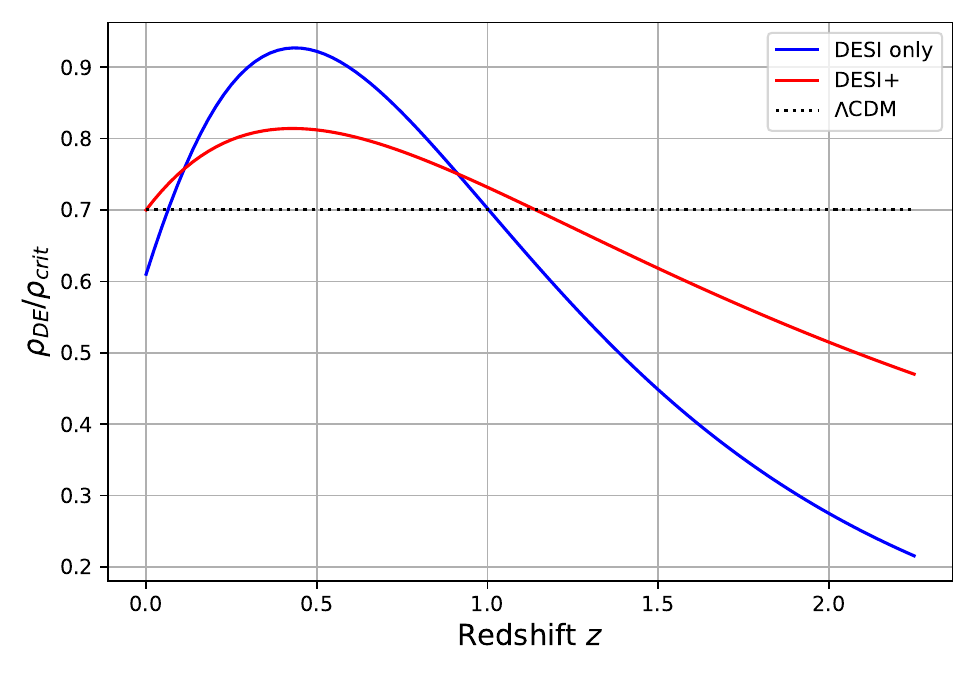}
\caption{Evolution of the energy density of dark energy for the DESI only best model ($\Omega_M = 0.39$, $w_0=-0.18$, $w_a = -2.73$), the DESI+ model ($\Omega_M = 0.30$, $w_0=-0.7$, $w_a = -1$), and $\Lambda$CDM ($\Omega_M = 0.3$).  
}
\label{rhoDE}     
\end{figure}

For the DESI+ model, $w$ increases from $-1.7$ at high redshift, crosses the phantom divide ($w=-1$), and asymptotically approaches large positive values in the future.  Further, $\rho_{DE}$  achieves a maximum around $a = \alpha /w_a \simeq 2/3 $, or $z\simeq 1/2$, where $w=-1$, and decreases rapidly earlier and later, shown in Fig.~\ref{rhoDE}.  

Considering the DESI BAO measurements alone, the $w_0w_a$ likelihood ellipse does not close and has a degeneracy direction, $w_a \simeq -3(1+w_0)$ or $\alpha \simeq 2w_a/3$, cf.~Fig.~6 in Ref.~\cite{DESI} and Fig.~11 in Ref.~\cite{DR2}.  This means the DESI data tightly constrain the position of the peak in $\rho_{DE}$, at $a = \alpha /w_a \simeq 2/3$ or $z \simeq 0.5$, but not the value of $w_a$, which determines the sharpness of the peak.  Adding CMB and SN data closes and narrows the ellipse, but does not change the degeneracy direction, cf. Fig.~11 in Ref.~\cite{DR2}.  

Redshift $z\simeq 0.5$ has other significance: it is the pivot point for DESI; namely, the redshift at which the DESI data have the most constraining power on $w$.\footnote{The pivot point for dark energy probes, first discussed in Ref.~\cite{DETF}, for BAO is around $z\approx 0.5$ and depends slightly upon the which other datasets it is combined with \cite{DR2}. }  As discussed in \S\ref{w0wa_revisited}, the only way that a $w_0w_a$ model can achieve $w=-1$ and $w_a \not= 0$ is at a minimum or maximum of $\rho_{DE}$.

While $w_0w_a$ has become a standard parameterization, it has inherent limitations, discussed next.  After that, we explore physics-inspired scalar-field models which can achieve a similar behavior for the expansion history preferred by DESI but without the odd asymptotic behavior.  We move on to a critical discussion of the DESI, CMB, and SN results and then end with a summary and some concluding remarks.

\section{The $w_0w_a$ phase space} \label{w0wa_revisited}

The $w_0w_a$ parameterization was designed and is used to test whether or not dark energy deviates from $\Lambda$. It is not a natural way to describe physically-based models of dark-energy, and it has significant weaknesses in modelling dark energy, especially scalar-field models. 

To better understand its limitations, we now discuss the kind of models that $w_0w_a$ can describe.  Recall that $\rho_{DE} \propto a^{-3\alpha} \exp [3w_a(a-1)], $
where $\alpha \equiv 1+w_0+w_a$.  From this, it follows that the generic behavior of dark energy described by $w_0w_a$ depends upon which quadrant the two parameters $\alpha$ and $w_a$  occupy:
\begin{enumerate}
\item SW ($\alpha , w_a <0$): $\rho_{DE}$ achieves a maximum for $a = \alpha / w_a$ where $w=-1$ and the phantom divide is crossed
\item NE ($\alpha , w_a >0$):  $\rho_{DE}$ achieves a minimum for $a = \alpha / w_a$ where $w=-1$ and the phantom divide is crossed
\item SE ($\alpha <0$, $w_a> 0$): $\rho_{DE}$ monotonically increases with $a$ and $w \le -1$ always
\item NW ($\alpha >0$, $w_a < 0$): $\rho_{DE}$ monotonically decreases with $a$ and $w \ge -1$ always
\end{enumerate}

\begin{figure}[tbp]
\center\includegraphics[width = 0.85\textwidth]{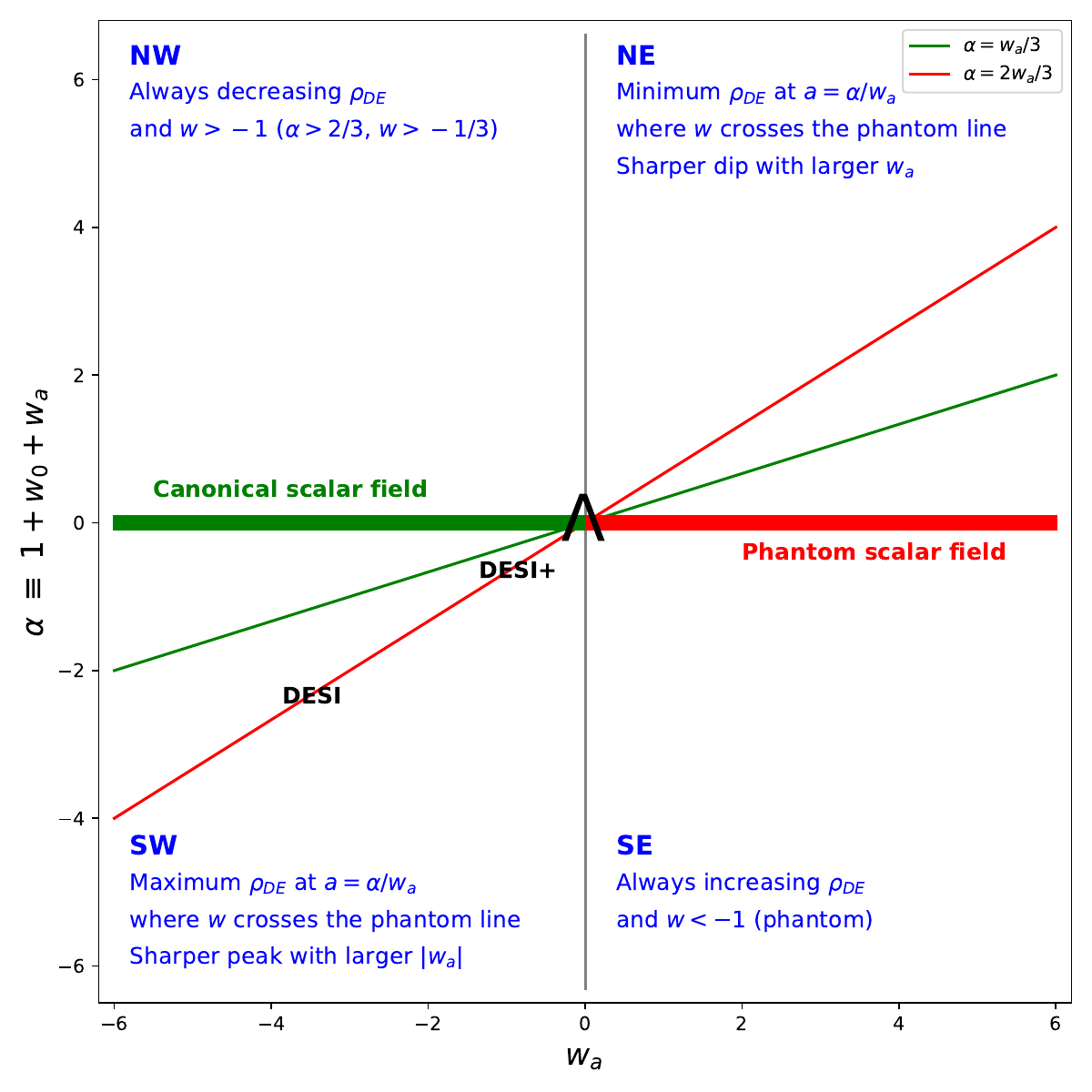}
\caption{The behavior of $w_0w_a$ models depends upon which quadrant the parameters $\alpha$ and $w_a$ lie in.  DESI and DESI+ mark the values preferred by DESI alone and DESI + additional datasets, respectively.  The red line corresponds to the DESI degeneracy direction.}
\label{quads}     
\end{figure}

\begin{figure}[tbp]
\center\includegraphics[width = 0.65\textwidth]{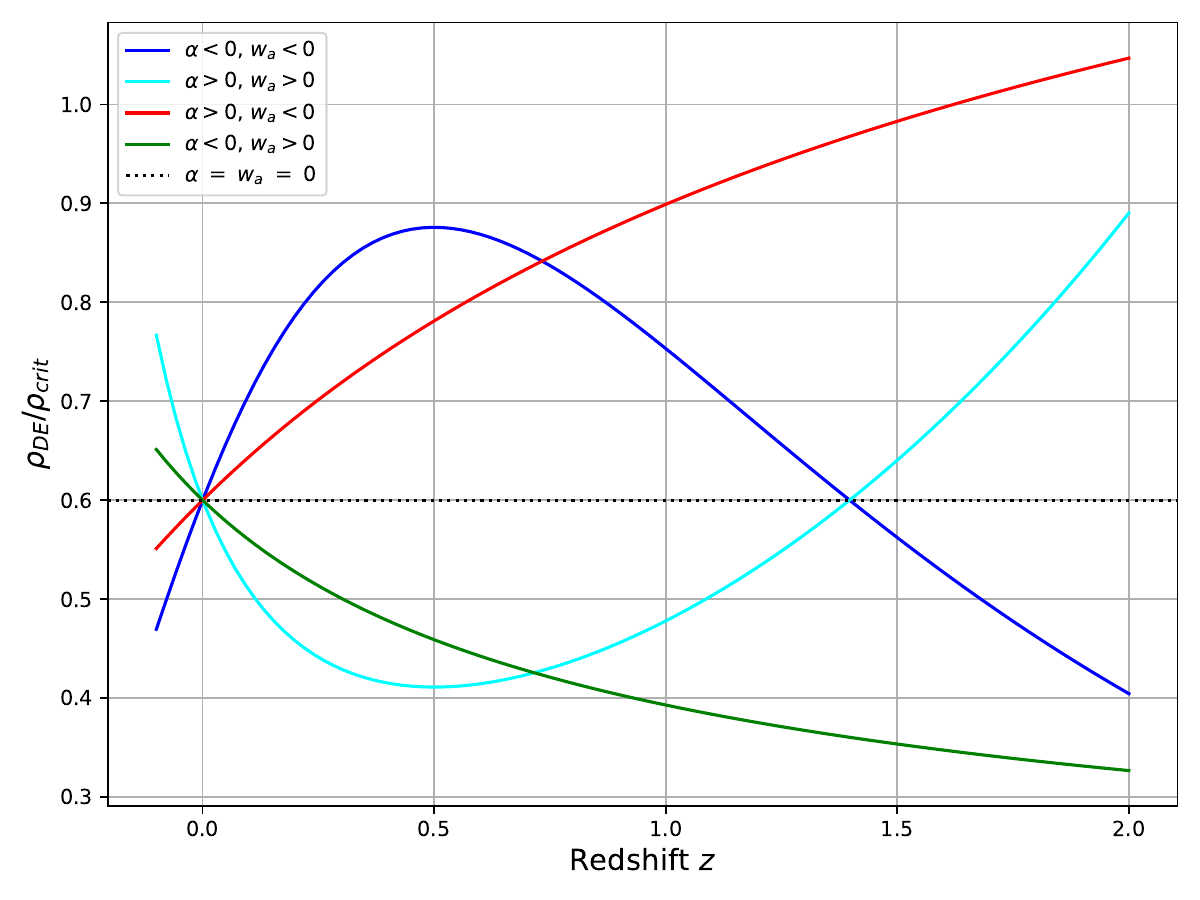}
\caption{The $w_0w_a$ parameterized energy density in the four quadrants: NE ($\alpha ,w_a >0$) $\rho_{DE}$ has a minimum (cyan);  SW ($\alpha ,w_a <0$) $\rho_{DE}$ has a maximum (blue), preferred by DESI; NW ($\alpha>0 ,w_a <0$) $\rho_{DE}$ decreases with $a$ (red), like a scalar field; NE ($\alpha<0, w_a >0$) $\rho_{DE}$ increases with $a$ (green), like a phantom field.  $\Lambda$ ($\alpha = w_a = 0$).}
\label{rhow0wa}     
\end{figure}

Only four types of dark energy evolution can be accommodated, and in half of the $\alpha$ -- $w_a$ plane, the NE and SW quadrants, $\rho_{DE}$ has unphysical behavior, crossing the phantom divide and achieving a minimum or maximum.  The $\alpha$-$w_a$ plane is summarized in Fig.~\ref{quads}, and the generic behaviors for $\rho_{DE}$ are shown in Fig.~\ref{rhow0wa}.

$\Lambda$ corresponds to the singular case, $w_a = \alpha = 0$, where $\rho_{DE} \propto \  const$.  A rolling scalar field with canonical kinetic energy corresponds to $\alpha = 0$ and $w_a <0$, and a phantom scalar field (negative kinetic energy term) corresponds to $\alpha =0 $ and $w_a >0$.  In both cases, $\rho_{DE}$ evolves as $\exp (3w_a a)$. Lastly, $w=-1$ can be achieved in only three ways:  $\alpha = w_a =0$ ($\equiv \Lambda$), or at the maximum ($w_0,w_a <0$) or minimum ($w_0, w_a >0$) of $\rho_{DE}$, where $w(z = w_a/\alpha - 1) = -1$.  

\begin{figure}[tbp]
\center\includegraphics[width = 0.75\textwidth]{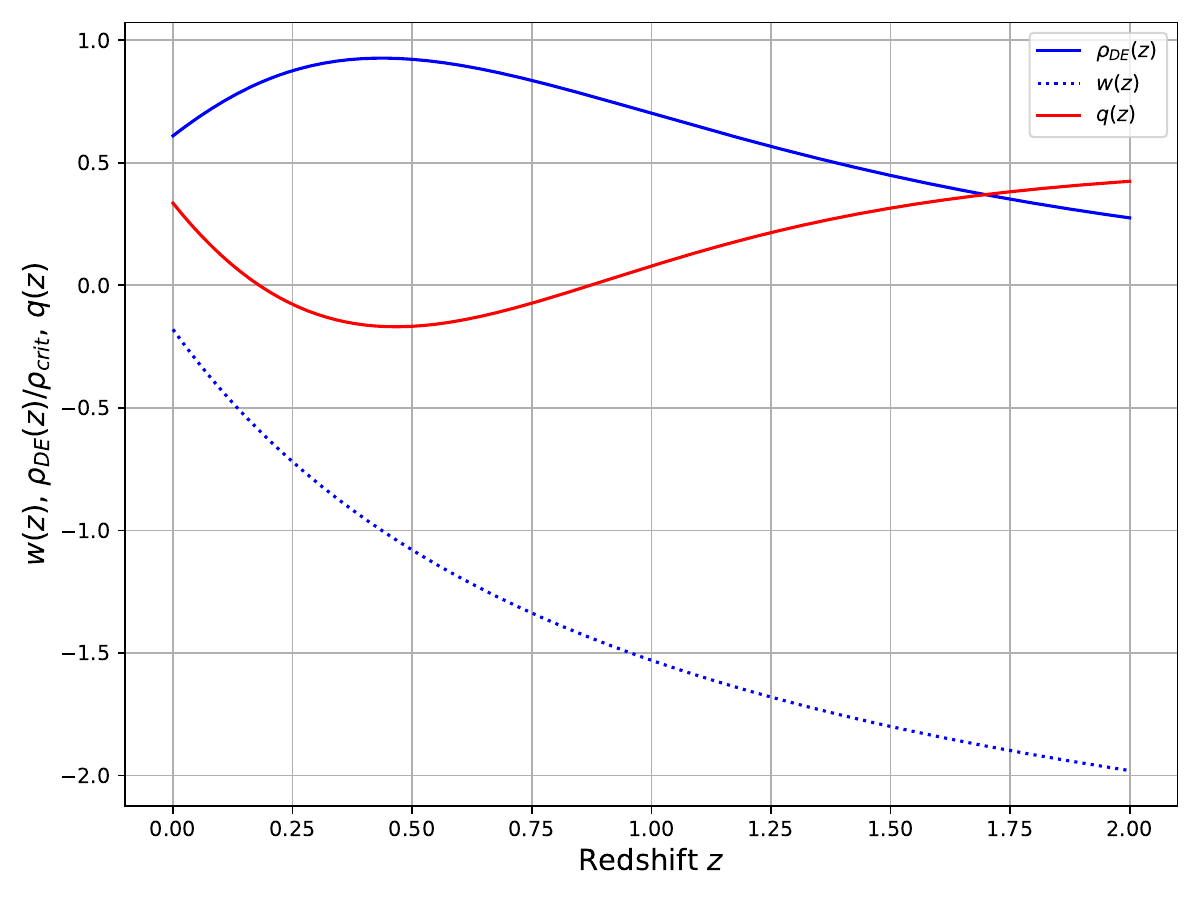}
\caption{Dark energy EOS and energy density, and the deceleration parameter for DESI-only best $w_0w_a$ model ($\Omega_M = 0.39$, $w_0 = -0.18$, $w_a = -2.73$ and $\alpha = -2.7$). Note, while this model accelerated in the past, but is not accelerating today.}
\label{EOSq}     
\end{figure}

The DESI-preferred $w_0w_a$ models -- with and without CMB and SN results -- are characterized $\alpha /w_a \simeq 2/3$ (and $w_a,\alpha <0$) so that $\rho_{DE}$ achieves a maximum at $z\simeq 0.5$ where $w = -1$.   
Larger $|w_a|$ corresponds to a sharper peak or trough, cf., Fig.~\ref{rhoDE}.  For the DESI-only model shown in Fig.~\ref{EOSq}, the falloff of $\rho_{DE}$ is so dramatic that the deceleration parameter today $q_0 \simeq 0.3$—that is, the Universe is not accelerating today, though it did so in the past.

\subsection{What drives the DESI models?}\label{pivotpointdrive}

The redshift $z\simeq 0.5$, where $\rho_{DE}$ peaks and $w=-1$ for the DESI-preferred models, is also the pivot point for the DESI data.  The question then arises:  is the need to have $w=-1$ at the pivot point driving the best-fit models, or is it the need to have a peaked dark energy?  Three clues point to the latter.  
First, the model preferred by the DESI results alone ($w_a = -2.73$)  is more sharply peaked than the DESI+ model ($w_a = -0.7$) which includes other datasets. Second, the DESI binned analysis of $w(z)$, cf. Fig.~12 in Ref.~\cite{DR2}, suggests that the DESI data want $w$ slightly more positive than $-1$ ($w\simeq -0.9\pm 0.05$) in the lowest redshift bin ($z\le 0.7$),  and more negative than $-1$ at higher redshift ($z>1$). Last, while a $w_0w_a$ model could have $w \simeq -1$ at $z\ge 0.5$ and $w > -1$ more recently (e.g., $\alpha = 0$ and $w_a = -0.3$, cf. Fig.~\ref{EOSw0wa}), the data prefer a maximum at $z \simeq 0.5$.

\section{Scalar-field dark energy}
A scalar field displaced from the minimum of its potential which is initially stuck due to Hubble friction is a model for evolving dark energy \cite{FHSW,PR}.  When the ``thawing''  field begins to roll, $w$ increases from $w= - 1$, at a rate determined by the shape of the potential.  The equations for the evolution of the cosmic scale factor $a(t)$ and  scalar field $\phi$ are:
\begin{eqnarray}
H^2 \equiv (\dot a/a)^2 & = & {8\pi (\rho_M + \rho_\phi  )\over 3 m_{\rm pl}^2} \\
\ddot\phi & = & - 3H\dot \phi - \partial V/\partial \phi \\
\rho_\phi & = & {1\over 2}\dot\phi^2 + V (\phi ) \\
p_\phi & = & {1\over 2}\dot\phi^2 - V (\phi )
\end{eqnarray}
where the EOS $w = p_\phi /\rho_\phi $, $\hbar = c =1$, $G=1/m_{\rm pl}^2$, dot denotes differentiation with respect to cosmic time ($d/dt$), and a spatially-flat Universe is assumed. 

\subsection{Three potentials}
We will consider three different scalar-field potentials, beginning with a massive scalar field with $V(\phi ) = {1\over 2}m^2\phi^2$. Its dimensionless equations-of-motion are:
\begin{eqnarray}
\theta^{\prime\prime} & = & -3(H/H_0) \theta^\prime - \beta \theta \\
a^\prime /a = H/H_0 &=& \left[ \Omega_M a^{-3} + {4\pi \over 3} \left({\theta^\prime}^2 + \beta \theta^2\right)\right] ^{1/2} \\
\rho_\phi & = & H_0^2 m_{\rm pl}^2 \left[ {1\over 2} {\theta^\prime}^2  + {1\over 2}\beta \theta^2 \right] \\
p_\phi & = & H_0^2 m_{\rm pl}^2 \left[ {1\over 2} {\theta^\prime}^2  - {1\over 2}\beta \theta^2 \right] 
\end{eqnarray}
where $\theta \equiv \phi /m_{\rm pl}$, $\beta \equiv m^2/H_0^2$, $\tau \equiv H_0t$, and prime denotes $d/d\tau$.  The dimensionless parameter $\beta$ controls how fast $\theta$ rolls, and $\beta \theta_i^2$ sets the initial energy density of the scalar field.  In the limit  $\beta \rightarrow 0$, $\Lambda$CDM is recovered.  With $a=1$ today, the matter energy density is $\rho_M = 3 m_{\rm pl}^2\Omega_MH_0^2 a^{-3}/8\pi,$ where $\Omega_M$ 
is the fraction of critical density contributed by matter (baryons + CDM).

A scalar field model has two parameters, $\Omega_M$ and $\beta$, compared to the three for a $w_0w_a$ model.  The initial value for $\theta$ is fixed by the requirement that when $a=1$ (today), $H/H_0$ should also equal one.  In the limit that $\beta \ll 1$, where $\phi$ evolves little, $\beta \theta_i^2 \simeq 3(1-\Omega_M)/4\pi,$ which we use as the initial guess for $\theta_i$.

We consider two other scalar field potentials:  a massless, self-interacting scalar field, $V( \phi ) = \lambda \phi^4/4$, and an exponential potential $V( \phi ) = V_0 e^{-\beta \phi / m_{pl}}$, motivated by the moduli fields of string theory.  The dimensionless equations-of-motions for the quartic potential are:
\begin{eqnarray}
\theta^{\prime\prime} & = & - 3(H/H_0) \theta^\prime - \beta\theta^3  \\
a^\prime /a &= & \left[ \Omega_M a^{-3} + {8\pi \over 3} \left( {1\over 2} {\theta^\prime}^2 + {1\over 4}\beta \theta^4 \right)\right] ^{1/2} \\
p_\phi & = & H_0^2 m_{\rm pl}^2 \left[ {1\over 2} {\theta^\prime}^2  - {1\over 4}\beta\theta^4 \right] \\
\rho_\phi & = & H_0^2 m_{\rm pl}^2 \left[ {1\over 2} {\theta^\prime}^2  + {1\over 4}\beta\theta^4 \right] 
\end{eqnarray}
where here the dimensionless parameter $\beta \equiv \lambda m_{pl}^2 /H_0^2 $.  

Like the massive scalar field model, there are two parameters, $\beta$ and $\Omega_M$, with $\theta_i$ being fixed by requiring that $H/H_0 =1$ when $a=1$. Again, in the limit of $\beta \rightarrow 0$, this model reduces to $\Lambda$CDM, and for small $\beta$, $\beta \theta_i^4 \simeq {3}(1-\Omega_M )/2\pi.$

The dimensionless equations-of-motions for the exponential potential are:
\begin{eqnarray}
\theta^{\prime\prime} & = & -3(H/H_0) \theta^\prime + \gamma \beta e^{-\beta \theta} \\
a^\prime /a &= & \left[ \Omega_M a^{-3} + {8\pi \over 3} \left( {1\over 2} {\theta^\prime}^2 + \gamma e^{-\beta \theta }\right)\right] ^{1/2} \\
p_\phi & = & H_0^2 m_{\rm pl}^2 \left[ {1\over 2} {\theta^\prime}^2  - \gamma e^{-\beta \theta} \right] \\
\rho_\phi & = & H_0^2 m_{\rm pl}^2 \left[ {1\over 2} {\theta^\prime}^2  + \gamma e^{-\beta \theta} \right]
\end{eqnarray}
where $\theta_i =0$ and the dimensionless parameter $\gamma \equiv V_0/H_0^2 m_{pl}^2$.  

As with the previous potentials, $\beta$ controls how fast the scalar field evolves.  This model too has two parameters, $\beta$ and $\Omega_M$, with $\gamma$ being fixed by the requirement that $H/H_0 =1$ when $a=1$.  Again, in the limit $\beta \rightarrow 0$, the model reduces to $\Lambda$, and for small $\beta$, $\gamma = {3} (1-\Omega_M)/8\pi.$

\subsection{Scalar fields and $w_0w_a$} \label{sf&w0wa}

How well can scalar field models be described by $w_0w_a$?  A scalar field begins with $w=-1$ at early times ($a\rightarrow 0$) when the field is stuck, and it increases (canonical kinetic term) or decreases (phantom) with the scale factor as it evolves.  So strictly speaking, the canonical scalar field maps onto the $\alpha = 0$ axis with $w_a <0$, and the phantom case maps onto the $\alpha = 0$ axis with $w_a >0$.  

For $\alpha =0$, the $w_0w_a$ solution for $\rho_{DE}$ depends only upon $w_a$,
$ \rho_{DE} \propto \exp (3w_a a)$.
For a canonical scalar field, the energy density decreases by a factor of $e^{3w_a}$ by the present; for a phantom field, it increases by the same factor. For $\beta = 3$, the fit is at the 10\% level because the exponential shape does not fit the evolution of a scalar field well, even over a restricted redshift range of $z=0$ to $z=4$.

\begin{figure}[tbp]
\center\includegraphics[width = 0.75\textwidth]{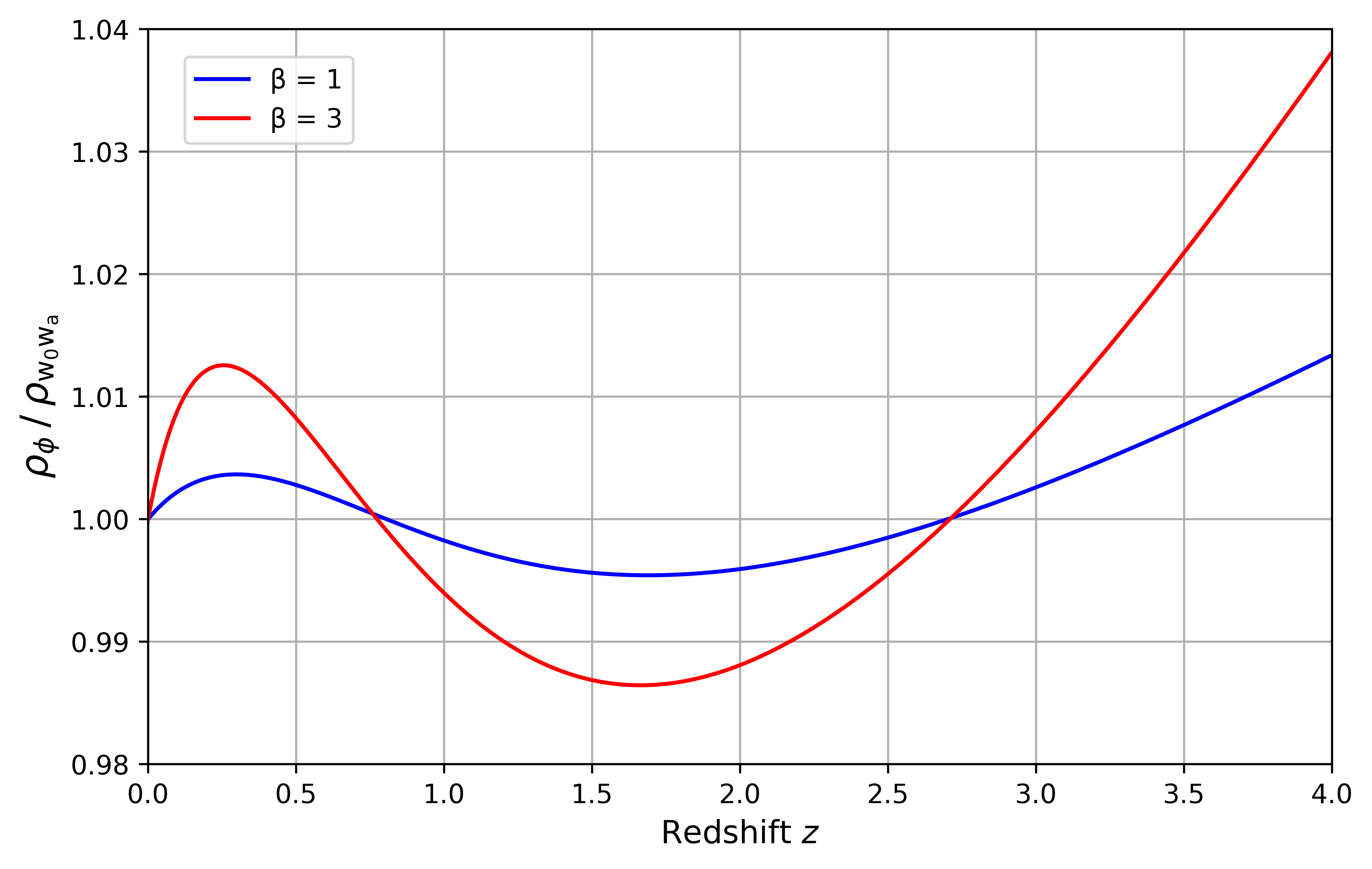}
\caption{Ratio of the energy density of a massive scalar field vs. $z$, with $\beta = 1$ (blue) and $\beta = 3$ (red), to a $w_0w_a$ model with $\alpha =-0.055$ and $w_a = -0.190$ (blue), and $\alpha =-0.154$ and $w_a = -0.537$ (red).}
\label{precision}   
\end{figure}

In practice, a scalar field model need only be represented over a finite range; for DESI and other galaxy and SN surveys, $z \simeq 0 - 4$ suffices.  Fig.~\ref{precision}  illustrates this point for the case of a massive scalar field:  for suitable parameters near $\alpha = 0$, the energy density of a scalar field can be approximated at the percent level for $z = 0 - 4$ for $\beta = 1$.   As $\beta$ increases, this becomes more difficult:  for $\beta = 3$, the precision drops to a few percent, which is not good enough for the DESI results whose precision approaches 0.3\%.  More complicated scalar potentials—especially if they have features—will be even more difficult to faithfully approximate.\footnote{This is at odds with statements made in Ref.~\cite{DESI}, just above Eq.~(5.4).  Also see \cite{RJS} for a discussion of the shortcomings of representing scalar-field dark energy by a $w_0w_a$ model.}

\subsection{Universal behaviour} \label{universal_sf_behavior}
Since $w_0w_a$ does not provide a robust way to represent scalar field models, we explore another method of representing different potentials.
We show that different scalar potentials can be made equivalent by a single parameter:  the initial slope of the potential.  

The initial slope of the scalar-field potential controls how fast the field begins to roll and hence how much the energy density deviates from a constant.  For small $\beta$, the field evolves very little until the present, and the initial slope of the potential captures much of its evolution. The dimensionless initial slope follows from the equations of motion:
\begin{eqnarray}
{\tilde V^\prime}_i & = & \beta \theta_i \simeq \beta^{1/2} (3\Omega_{DE}/4\pi )^{1/2} \qquad {\rm massive} \nonumber \\
{\tilde V^\prime}_i & = & \beta \theta_i^3 \simeq \beta^{1/4} (3\Omega_{DE}/2\pi )^{3/4} \qquad {\rm quartic} \nonumber \\
{\tilde V^\prime}_i & = & \gamma {\tilde V}_i \simeq \beta (3\Omega_{DE}/8\pi ) \qquad \qquad \ \  {\rm exponential} \nonumber 
\end{eqnarray}
where the second expressions follow by substituting the $\beta \rightarrow 0$ expression for $\theta_i$ and $\gamma$ and are approximate.

\begin{figure}[tbp]
\center\includegraphics[width =0.75\textwidth]{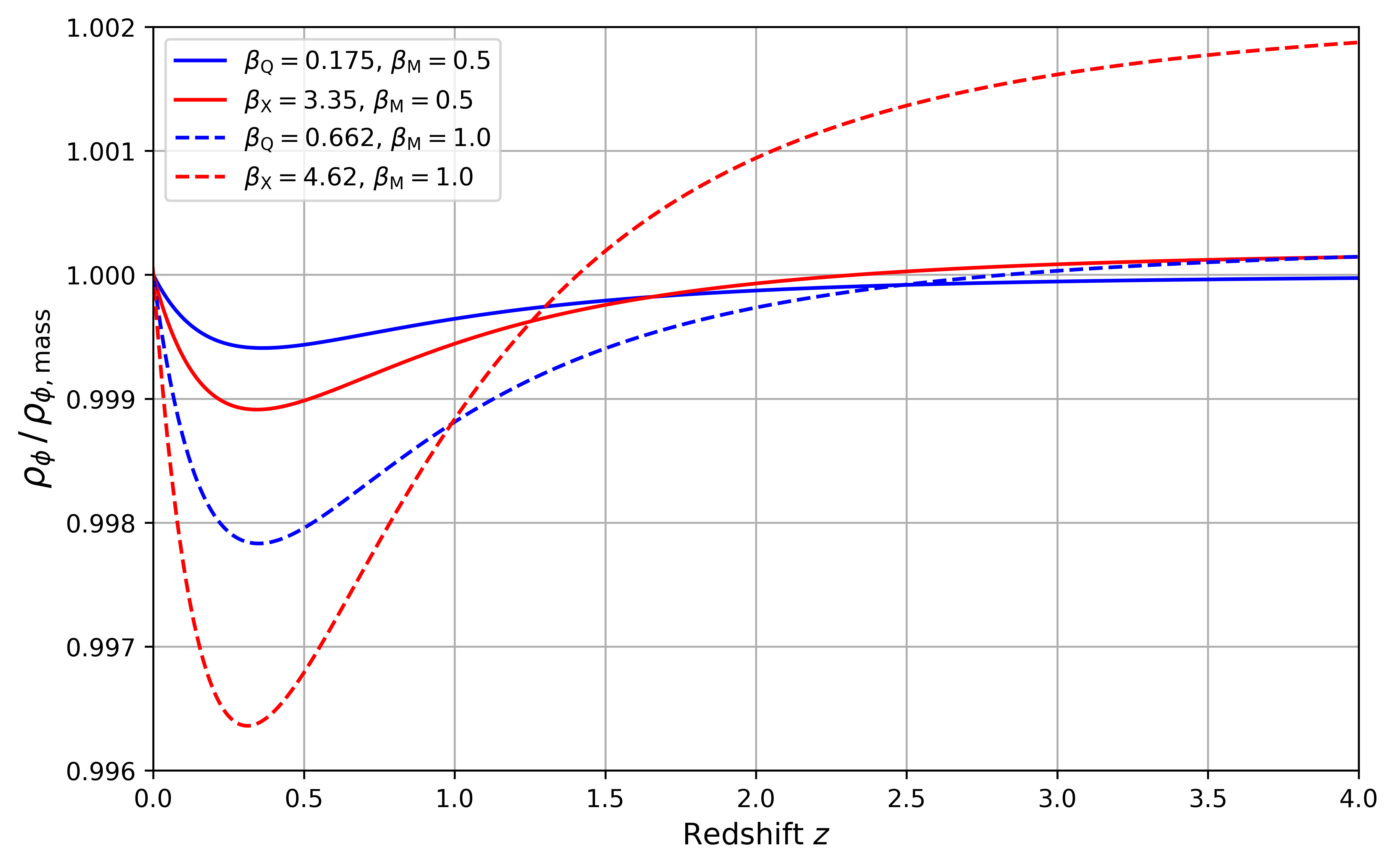}
\caption{$\rho_{DE}$ for quartic (blue) and exponential (red) potentials compared to a massive scalar field with $\beta_{M} = 0.5, 1$.  Note, $\beta_Q, \beta_X$ are slightly smaller than the values estimated in Eq.~\ref{equivalentbeta}, $0.187, 0.748$ for the quartic potential and $3.46,4.89$ for the exponential potential.}
\label{universal}   
\end{figure}

The quartic and exponential models can be mapped onto the massive scalar-field model by solving for the values of $\beta$ that give the same dimensionless slope:
\begin{eqnarray}
\beta_{Q}& \simeq & \left({\pi \over 6\Omega_{DE}}\right)\beta_{Meq}^2  \qquad {\rm quartic} \nonumber \\
\beta_{X}& \simeq &  \left({16\pi \over 3\Omega_{DE}}\right)^{1/2} \beta_{Meq}^{1/2} \qquad \ \   {\rm exponential}
\label{equivalentbeta}
\end{eqnarray}
Using these approximate formulae that match a massive scalar field with a quartic or exponential potential, we find that for $\beta_M \le 1$ and $z \le 4$, the agreement between the scalar field energy density for the quartic and exponential potentials with their matched massive scalar field is better than 3\%.  Because the matching formulae are only approximate, we can adjust $\beta$ to improve the agreement. After doing so, the agreement for $\beta_M \le 1$ for $z\le 4$ improves to better than 0.25\%, cf.~Fig.~\ref{universal}.  

In sum, a massive scalar field can represent the energy density of a scalar field with a quartic or exponential potential to sub-percent precision.  Further, since dark energy is only a fraction of the total energy density, the precision is even greater for the expansion rate, $\delta H(z)/H(z) \simeq 0.35 (\delta \rho_{DE}/\rho_{DE})/[0.7 + 0.3(1+z)^3] \le 0.1$\%.

\section{Varying dark energy confronts the data}\label{comparingtoDESI}
In this section we critically compare dark energy models with the data to assess the strength of the evidence for evolving dark energy.  Rather extracting parameters by combining all the relevant data, we examine what the DESI, CMB and SN data imply individually and in combination for $\Lambda$CDM, $w_0w_a$ models and scalar field dark energy.

\subsection{DESI BAO distances}

DESI uses BAO to measure the following cosmic distances at redshifts $z = 0.295,$ $ 0.510, 0.706,$ $ 0.930,$ $ 1.317,$ $ 1.491,$ and $ 2.330$.\footnote{In Table 1 of version 3 of the DR1 paper, they have added a {\it fourth} significant figure to the effective redshift; this makes a significant difference in comparing the data with models.} 
\begin{eqnarray}
D_M(z) & \equiv & r(z) = \int_0^z {dz \over H(z)} \\
D_H(z) & \equiv & {1 \over H(z)} \\
D_V(z) & \equiv & \left[ z D_M(z)^2 D_H(z) \right]^{1/3}
\end{eqnarray}
We have computed $\chi^2$ for the 13 measured distances in DR2\footnote{The formula for the 12 DR1 distance measurements is analogous, with two $D_V$, five $D_H$, and five $D_M$ data points, cf.~Table 1 in Ref.~\cite{DESI}.}: 
$$\chi^2 = {(\Delta_1 D_V)^2 \over \sigma^2_1 (D_V)}+\sum_{i = 2}^7 {1\over 1-r_i^2} \left[ { (\Delta_i D_H)^2 \over \sigma^2_i(D_H)} + { (\Delta_i D_M)^2 \over \sigma^2_i(D_M)} -{2r_i \Delta_i D_H\Delta_i D_M \over \sigma_i(D_H)\sigma_i(D_M)}\right] $$
where $\Delta_i D_H \equiv D_H({\rm theory}) - D_H ({\rm data})$ for data point i in Table IV \cite{DR2}, with $\Delta_i D_M$ and $\Delta_i D_V$ defined similarly. The $D_M$ and $D_H$ measurements are connected by correlation coefficients $r_i$, given along with the data.

For a quantitative comparison between models that takes model complexity into account, we use the penalty-based Bayesian Information Criterion (BIC):
$$ {\rm BIC} = \chi^2 + k\ln(n) , $$
where $k$ is the number of parameters in the model and $n$ the number of observations in the dataset ($n=13$ for DR2).
For $\Lambda$CDM, $k=2$ ($H_0r_d$ and $\Omega_M$); the scalar-field models add one extra parameter $\beta$ ($k=3$), and two parameters are added for $w_0w_a$ models ($k=4$). 

\subsection{$\Lambda$CDM, $w_0w_a$ and scalar fields vs. DESI} \label{Lw0wa}
The DESI BAO distances are expressed in units of the drag horizon, 
\begin{eqnarray} 
r_d \equiv \int_{z_d}^\infty {v_s (z) dz \over H(z)} = \sqrt{{1\over 3}}  \int_{z_d}^\infty {dz \over H(z)\sqrt{1+c/(1+z)}},
\label{drag_horizon}
\end{eqnarray}
where $z_d \simeq 1060$, $v_s$ is the sound speed, and $c \equiv {3\over 4} \Omega_Bh^2/\Omega_\gamma h^2 \simeq 655 \pm 3$, which is fixed by the BBN determination\footnote{The CMB determination by Planck is consistent with the BBN value and has a similar uncertainty.} of $\Omega_B h^2 = 0.0224 \pm 0.0002$ \cite{Cooke}, and the COBE measurement of the temperature of the CMB, $T_0 = 2.7255\pm 0.0006\,$K \cite{Fixsen}.  

Theoretical models predict distances in units of $H_0^{-1}$, and so the dimensionless parameter $H_0r_d$ is the ``link'' needed to compare models with the DESI data.  We note that $H_0r_d$ also ties together cosmology at low redshift (through $H_0$) and high redshift ($z > r_d \simeq 1060$).  
We consider two ways to fix $H_0r_d$.  The first is to be agnostic about the physics at high redshift, treating $H_0r_d$ as another parameter to determine or marginalize over.  The second is to take advantage of the power of the CMB to precisely fix $H_0r_d$, cf. \S\ref{CMB_constraint}.

For $\Lambda$CDM, we find that $\chi^2$ is minimized at a value of $10.5$ for:\footnote{This agrees with Eq.17 of Ref.~\cite{DR2}, taking into the fact that we have set $c=1$.} $\Omega_M = 0.297$ and $H_0r_d = 3.388 \times 10^{-2}$.
For the $w_0w_a$ models we find that $\chi^2$ is minimized at a value of $5.69$ for $\Omega_M  =  0.39$, $H_0r_d  =  3.039 \times 10^{-2}$, $w_0  =  -0.18$, and $w_a  =  -2.73 $. Compared to $\Lambda$CDM, $\Delta \chi^2 = -4.8$. However, the BIC value for $\Lambda$CDM is still lower, 15.6 vs. 15.9.

Note, we adopted a uniform prior of $\Omega_M=0.2-0.4$ because the DESI data alone do not break the degeneracy between $\Omega_M$, $w_0$, and $w_a$, resulting in unclosed contours in the $w_0$-$w_a$ parameter plane, cf. Fig.~11 in Ref.~\cite{DR2}. In particular, $\chi^2$ decreases with decreasing $w_a$ at the expense of increasing $\Omega_M$.  This aforementioned $w_0w_a$ model with $\Omega_M \approx 0.4$ is inconsistent with a host of other data.

Moving on to the massive scalar field models; as discussed in \S\ref{universal_sf_behavior}, they are a stand-in for all the scalar field models. As shown in Fig.~\ref{DR2Massive} a massive scalar field model with $\beta \simeq 0.40$ has a slightly lower $\chi^2$, though at the expense of a higher BIC value.  However, a prior for $\Omega_M$ is not needed to break the degeneracy between the model parameters, and $\chi^2$ is minimized for $\Omega_M \approx 0.30$, cf. Fig.~\ref{Contours}, which is more in line with other determinations of the matter density.

\subsection{CMB constraint to $H_0r_d$}\label{CMB_constraint}
Next, we take advantage of the close relationship between the drag horizon and sound horizon and the fact that the CMB precisely determines the sound horizon to determine $H_0r_d$.  In particular, the sound horizon 
$$H_0r_s = \sqrt{{1\over 3}} \int_{z_s}^\infty {H_0dz \over H(z)\sqrt{1+c/(1+z)}},$$
where $z_s \simeq 1090$.\footnote{A similar approach to fixing $H_0r_d$ is described in detail in Ref.~\cite{Kosowskyetal}}
Measurements of the position of the first peak in the CMB angular power spectrum, $\theta_* = 1.0411 \times 10^{-2} \pm 3\times 10^{-6} $ \cite{PlanckLegacy}, together with the comoving distance to the last-scattering surface, i.e., $r(z_s) = D_M(z_s)$, fix the sound horizon: $H_0 r_s  \equiv \theta_* H_0 r(z_s).$

The sound horizon and drag horizon only differ by a small amount:
\begin{eqnarray}
H_0r_d & =  & H_0r_s + \sqrt{{1\over 3}}  \int_{z_d}^{z_s} {H_0 dz \over H(z) \sqrt{1+c/(1+z)}},
\label{DragHorizon2}
\end{eqnarray}
Because the relevant redshifts are near matter-radiation equality, it is important to include radiation in the formula for the expansion rate:
$$H(z)/H_0 = \left[ \Omega_M (1+z)^3 + \Omega_R (1+z)^4 + {\rm DE}\right]^{1/2} $$
where $\Omega= \, _R = \Omega_\gamma [1 + 21(4/11)^{4/3}/8] \simeq 1.68 \Omega_\gamma \simeq 8.79 \times 10^{-5}$ (for $h=0.7$) and `DE' stands for the dark energy contribution.  For $\Lambda$CDM, `DE'\,=\, $1 - \Omega_M$.

The first term in Eq.~\ref{DragHorizon2}, $H_0r_s$, is a function of $\Omega_M$, the dark-energy model 
and $h$; it depends very weakly upon $h$, around $\pm 0.14\%$ for $h = 0.6 - 0.8$.  The integral term in Eq.~\ref{DragHorizon2} depends upon $\Omega_M$ and even more weakly upon $h$.  

Bringing it all together, for a given $\Omega_M$ and dark-energy parameters, the CMB measurement of $\theta_*$ fixes $H_0r_d$ very precisely, and $H_0r_d$  no longer needs to be determined by minimizing $\chi^2$ or marginalizing over.  Since $r_d$ itself is not fixed, we are in essence using BAO as an uncalibrated standard ruler.

The shortcomings of fixing $H_0r_d$ with the CMB constraint to $\theta_*$ are the possibility that dark energy is important around last scattering, e.g., early dark energy, and the fact that it depends upon the full cosmological history between $z =0$ and $z=z_s$. 

\subsection{CMB + DESI vs. models} \label{SFDE}

Using the CMB constraint, $\Lambda$CDM is still a good fit to the DESI DR2 distance measurements: $\chi^2 = 10.5$ and BIC value of $15.6$. However, the best $w_0w_a$ model is a significantly better fit for  $\Omega_M = 0.346$,  $w_0 =-0.499$ and  $w_a = -1.461$: $\Delta\chi^2 = -3.4$, but with a higher BIC, $\Delta\text{BIC}=+1.71$.  Implementing the CMB constraint has also broken the degeneracy between $\Omega_M$, $w_0$ and $w_a$ and resulted in a lower - and more favorable - value of $\Omega_M$.

\begin{figure}[tbp]
\center\includegraphics[width = 0.7\textwidth]{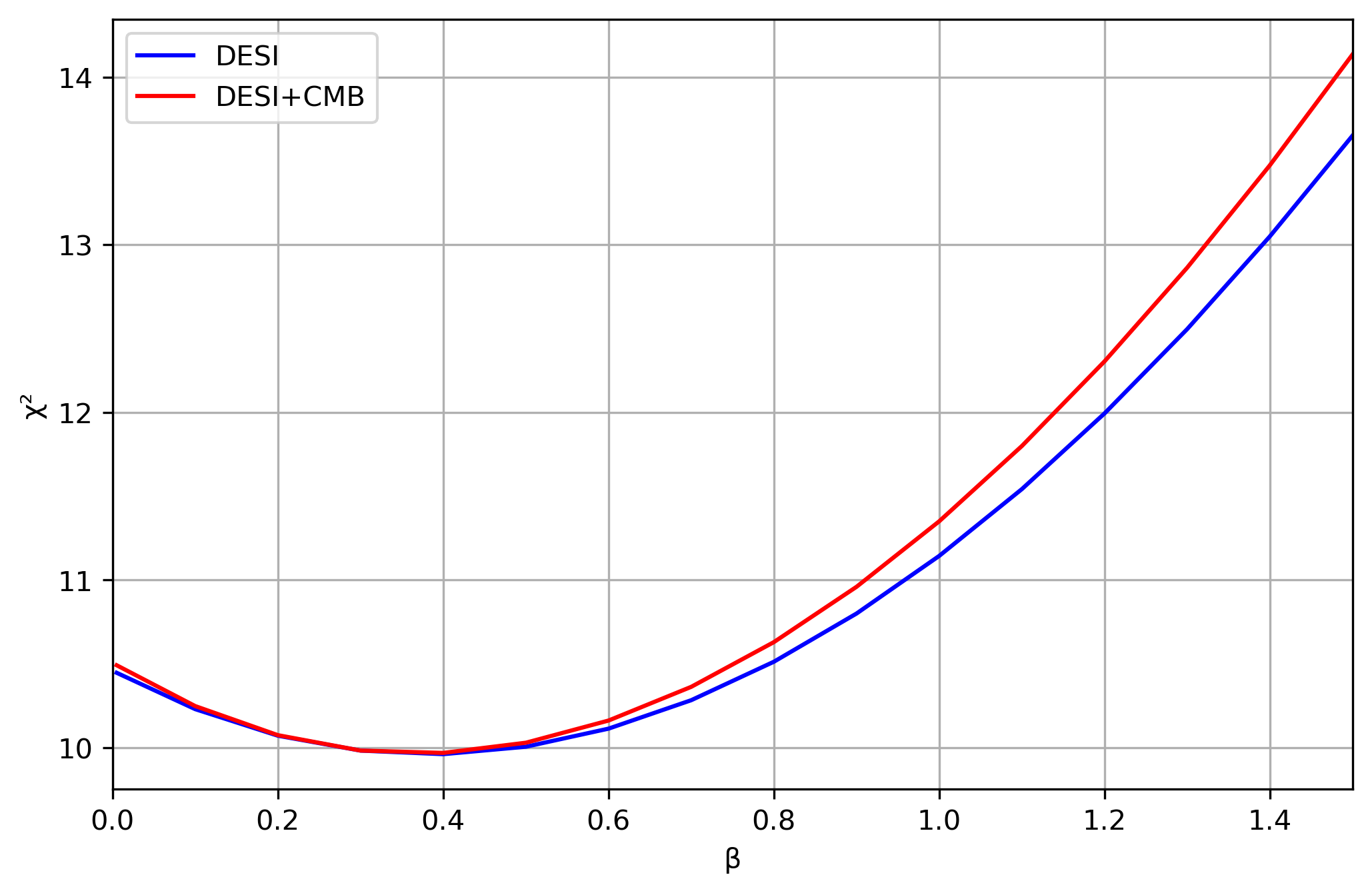}
\caption{$\chi^2$ vs. $\beta$ for the massive scalar field model and DESI with (and without) the CMB constraint, for the value of $\Omega_M$ (and $H_0r_d$) that minimizes $\chi^2$.  Note that the CMB constraint has very little impact on $\chi^2$}
\label{DR2Massive}     
\end{figure}

Fig.~\ref{DR2Massive} shows $\chi^2$ as a function of $\beta$ for the massive scalar field model with the CMB constraint.  It is minimized for $\beta \simeq 0.4$ at a value slightly lower than $\Lambda$CDM, $\Delta\chi^2 = -0.5$, with a corresponding BIC value of 20.2, which is higher than that of $\Lambda$CDM. 

We have focused on the full DESI (DR2) dataset; it is worth noting that they are consistent with a similar analysis of the first-year (DR1) results.
The larger DR2 dataset has similar preferred parameter values ($\Omega_M, w_0, w_a$) at a higher significance, and slightly stronger evidence for evolving dark energy \cite{DR2}.

\subsection{Pantheon+ SN}
Type Ia SN probe cosmological distances over a similar redshift range as DESI, with much less precision per measurement (typically 20\% or so), but with many more distance measurements and different systematic errors.  We have used the Pantheon+ data set of 1701 SN \cite{Pantheon}, which extends to redshift $z = 2.26$, to constrain dark energy models. 

The SN distances are expressed in units of $H_0^{-1}$.  The actual value of $H_0$ does not affect the dark energy analysis, and so $H_0$ is a nuisance parameter to be marginalized over.\footnote{We have checked this by adding 0.1 to the distance modulus of all supernovae-it does not affect the results.}  Like the CMB determination of $H_0r_d$, the SN are, for our purposes, an uncalibrated standard ruler.

\begin{figure}[tbp]
\center\includegraphics[width = 0.75\textwidth]{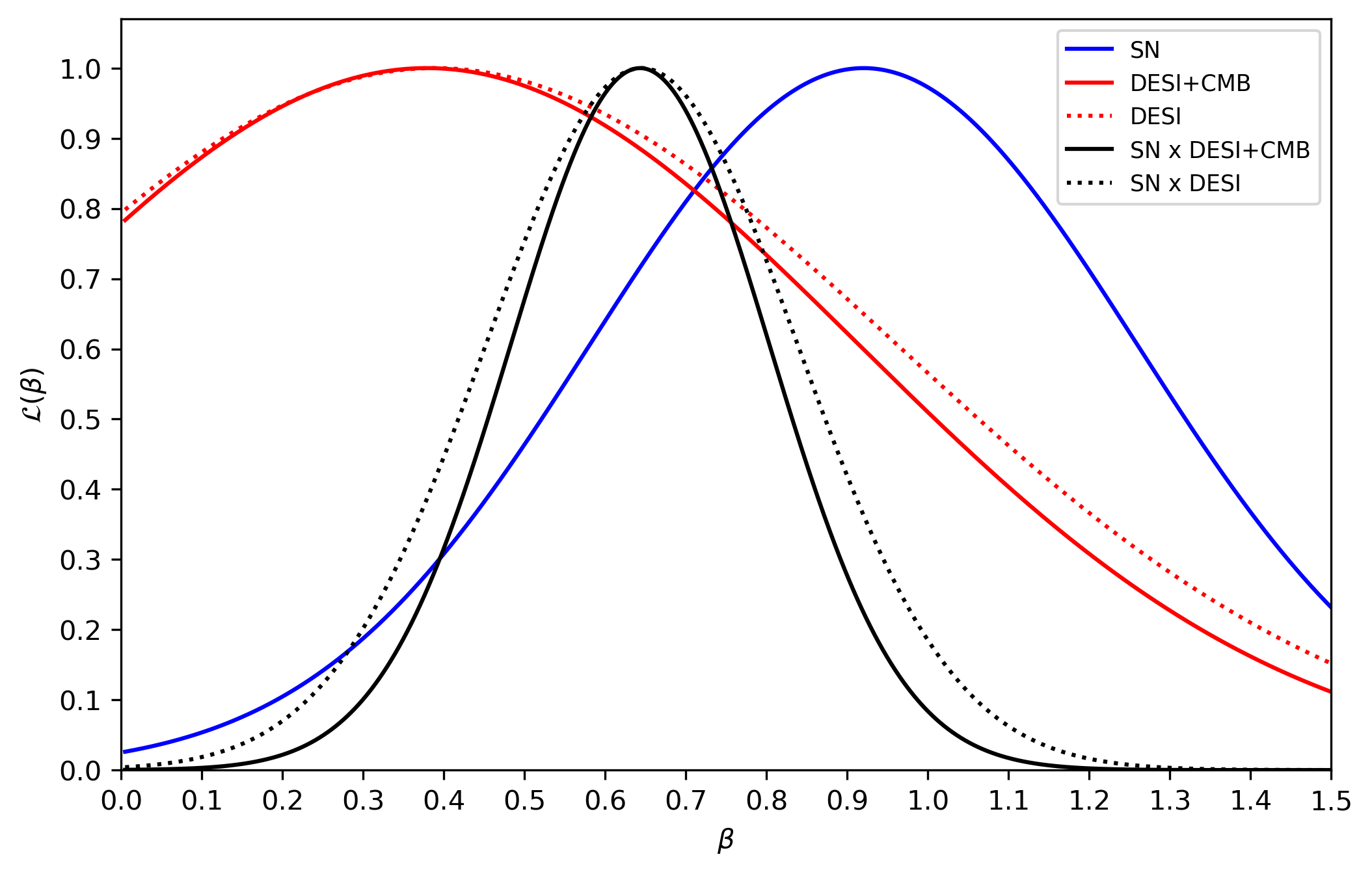}
\caption{Likelihoods marginalized over $\Omega_M$ for: DESI (with and without the CMB constraint), Pantheon+ SN, and the DESI+SN dataset. (Without the CMB, the likelihoods are also marginalized over $H_0r_d$.) The 95\% credible ranges are: $\beta = 0 - 1.30$ (DESI), $\beta = 0 - 1.24$ (DESI+CMB), $\beta = 0.35 - 1.49$ (SN), $\beta = 0.27 - 1.02$ (DESI+SN), and $\beta = 0.33 - 0.96$ (DESI+CMB+SN). (For the credible ranges a physical prior, $\beta >0$ was imposed.)}
\label{Likelihoods}     
\end{figure}
\begin{figure}[tbp]
\center\includegraphics[width = 0.85\textwidth]{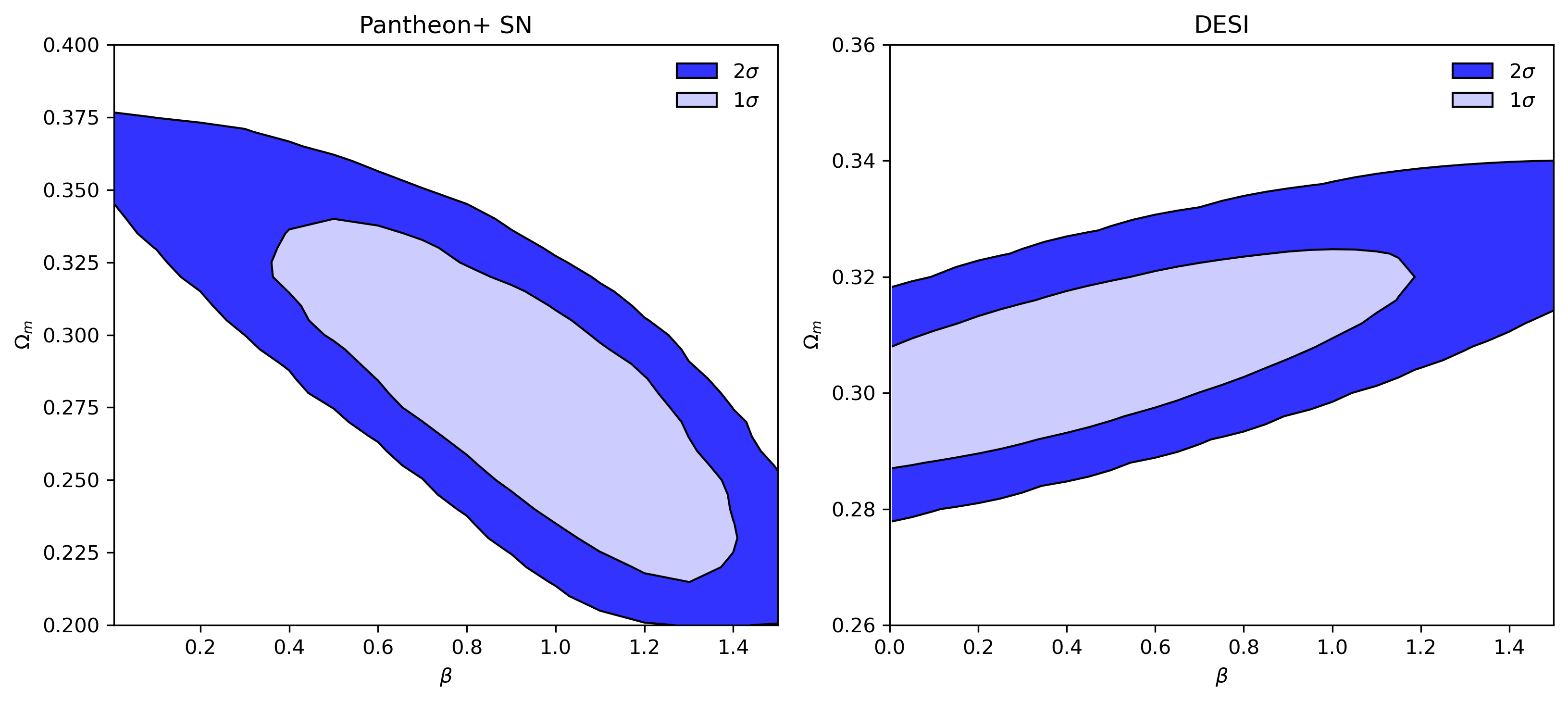}
\caption{Two-dimensional $\Omega_M$-$\beta$ likelihood $1\sigma$ and $2\sigma$ contours obtained from the Pantheon+ dataset and the DESI dataset.}
\label{Contours}     
\end{figure}

We have computed $\chi^2$ values using their covariance matrices, cf. \S2.2 and \S2.3 of Ref.~\cite{Pantheon}.  From this we have computed the likelihood function, $\mathcal{L} \propto \exp ({-\chi^2}/2)$, which is a function of $\Omega_M$ and $\beta$ for scalar field models.
We have also computed the likelihood function for the DR2 dataset as a function of $\Omega_M$ and $\beta$.  The results, marginalized over $\Omega_M$,  are shown in Fig.~\ref{Likelihoods}.  Fig.~\ref{Contours} shows the likelihood contours in the $\Omega_M - \beta$ plane for both DR2 and SN.

The SN results alone favor a value of $\beta \sim 1$, where $\Delta\chi^2$ decreases to a minimum of around $\Delta\chi^2 = -7$ compared to $\Lambda$CDM.  However, $\Lambda$CDM still is the best performing model with, $\Delta\text{BIC}=0.44$ for a scalar field, and the 95\% credible range is $\beta = 0.35 - 1.49$.  For the SN data alone, the best $w_0w_a$ model does not perform as well, $\Delta\chi^2 = - 5.8$ and $\Delta\text{BIC}=9.08$, for $\Omega_M = 0.4$, $w_0 = -0.72$ and $w_a = - 2.77$.  

The DR2 results alone weakly favor a lower value, $\beta \sim 0.4$, with the 95\% credible range $\beta = 0 - 1.30$.  Also shown in Fig.~\ref{Likelihoods} is the joint DESI+SN likelihood marginalized over $\Omega_M$. The 95\% credible range is  $\beta = 0.27 - 1.02$.

To summarize, the DESI results alone prefer a $w_0w_a$ model to a scalar field, while the SN data alone mildly prefer a rolling scalar field.  Further, for a scalar field, both DESI and the SN data prefer a lower value of $\Omega_M \simeq 0.3$, consistent with other determinations, than $w_0w_a$ models, which prefer a larger value, $\Omega_M \simeq 0.35$. Together, the CMB, DR2, and SN  favor scalar-field dark energy over a $w_0w_a$ model of evolving dark energy.

\subsection{Low-redshift data only}

Using the CMB to fix $H_0r_d$ ties the low-redshift DESI and SN results to the behaviour of $H(z)$ from $z\simeq 2 - 1090$, which could be impacted by new, unaccounted-for physics at these redshifts.  In addition, the $w_0w_a$ parameterization was not meant to be a physical model, and extending it to redshifts $z\gg 2$ could misrepresent the actual behavior of dark energy at these redshifts.  If instead, one marginalizes over $H_0r_d$, the DESI and SN results can be combined without any dependence upon  physics beyond redshift $z\simeq 2$.

The results of this approach are shown in Fig.~\ref{Likelihoods}.  They are qualitatively similar to those obtained using the CMB constraint:  the DESI curve is shifted to larger $\beta$ and the joint DESI+SN likelihood is almost unaffected.  In the context of scalar-field models, this speaks to the robustness of the evidence for evolving dark energy, as it does not depend significantly upon the evolution of $H(z)$ beyond $z\simeq 2$.

\section{Concluding remarks} \label{conclusions}

Before we summarize and provide some final remarks, we briefly mention an intriguing alternative explanation for the peaked dark energy that the DESI results prefer.

\subsection{Dark-energy bump or mirage?} \label{DE,bump,mirage}

The apparent strange evolution of dark energy can arise if the matter density does not evolve precisely as $a^{-3}$.  The key point is that DESI assumes a two-component universe — dark matter and dark energy — and infers $\rho_{DE}$ by subtraction.  

Specifically, DESI determines the total energy density $\rho_{T}$ from measurements of the expansion rate $H(a)$,
$ \rho_{T}(a) = { 3H^2(a) / 8\pi G},$ and then subtracts the matter component,
$\rho_M = {3H_0^2\Omega_M / 8\pi G a^3},$ assuming a flat Universe. 
Thus, the ``DESI'' inferred dark energy is
\begin{eqnarray}
\rho_{DE-SI} = \left[ {H^2\over H_0^2} - {\Omega_M\over a^3} \right] \rho_{crit} ,
\end{eqnarray}
where $\rho_{crit} \equiv 3H_0^2/8\pi G$ is the critical density today.

If the {actual} matter density does behave as $\rho_M \propto a^{-3}$, this procedure leads to a misleading dark-energy evolution.  Suppose that the mass $m$ of the dark matter particle evolves, $m = m(a)$, with an asymptotically constant value as $a\rightarrow 0$, $m(a) \rightarrow m_0$.\footnote{We make this assumption because any significant variation in mass of the dark matter particle while structure is forming is likely ruled out.}
With this setup, the dark-energy energy density inferred by DESI would be:
\begin{eqnarray}
\rho_{DE-SI}  =\left[ \rho_{DE} + \left( {m(a) \over m_0} -1 \right) {\Omega_M \over a^3} \rho_{crit} \right]  ,
\end{eqnarray}
where $\Omega_M$ is the fraction of critical density today had the mass of the dark matter particle not varied.  The inferred dark energy differs from the real dark energy by a term that involves the variation of the dark-matter particle's mass.
Without worrying about the mechanism for the variation 
or possible coupling between dark matter and dark energy, here is how a bump in the inferred dark-energy would arise:  the mass of the dark matter particle rises as the Universe expands and then levels off around $z\sim 0.5$, and then, the dark energy density decays, as it would for a rolling scalar field.  This leads to a rise in the inferred dark energy, $\rho_{DE-SI}$, until a redshift $z\simeq 0.5$, and then a fall thereafter.

The presence of any additional component to the energy density of the Universe beyond matter and dark energy could lead to a strange behavior inferred for the dark energy.

\subsection{Summary}

In this paper we examined the DESI results concerning the nature of the dark energy.  We focused on a better understanding of the $w_0w_a$ models that underpin their claims for evolving dark energy, explored physics-based models involving a scalar-field, and studied the interplay between the DESI results and CMB and SN data. What follows is a brief summary.

$\Lambda$CDM is a good fit to the DESI results and any combination with the CMB and SN results. Further, while $w_0w_a$ models and scalar field models can have a lower $\chi^2$ value, they have more parameters, and under the BIC the best-performing model is $\Lambda$CDM. 

Both the best $w_0w_a$ DESI-only model ($\Omega_M = 0.39$, $w_0=-0.18$, $w_a=-2.73$) and the DESI+ $w_0w_a$ model ($w_0=-0.7$, $w_a=-1$) describe a dark-energy component that achieves a maximum energy density around $z\simeq 0.5$, falling off rapidly earlier and later. This leads to a bump in $H(z)$ relative to $\Lambda$CDM of a few percent around $z\simeq 0.5$.

The favored ellipse in the $w_0w_a$ plane has a degeneracy direction, $w_0 \simeq -1 -w_a/3$.  This direction corresponds to the peak in the energy density of the dark energy around $z\approx 0.5$. Considering the DESI results only, the ellipse does not close; combining the DESI results with CMB and/or SN data closes and narrows the ellipse, cf. Fig.~11 in Ref.~\cite{DR2}, but does not change the degeneracy axis.

At the redshift of this peak, $w(z) = -1$, and DESI most tightly constrains $w$. This raises the question: Is it the need to have $w=-1$ at the pivot point or the need for dark energy to be peaked which drives this behaviour?  We find evidence for the latter, cf.~\S\ref{pivotpointdrive}.

The $w_0w_a$ parameterization has limited ability to model dark energy, with only four generic behaviors -- monotonically increasing or decreasing, or with a maximum or minimum -- determined by which quadrant $\alpha = 1+w_0+w_a$ and $w_a$ lie in.  Further, $w_0w_a$ is a one-dimensional parameterization of scalar field models, and unless $\beta$ is small, it cannot represent scalar field dark energy to the sub-percent level needed for the DESI results.

Our scalar-field dark energy models are described by a single additional parameter $\beta$, with $\beta \rightarrow 0$ reducing to $\Lambda$CDM.  
For $\beta \le 1$, a variety of scalar-field potentials can be very accurately represented by a massive scalar field.

While the DESI results favor a $w_0w_a$ model over a scalar-field model, the Pantheon+ SN results favor a massive scalar field model with $\beta \sim 1$ over a $w_0w_a$ model.  Further, for a scalar-field model, $\Omega_M \simeq 0.3$ rather than $0.35$ or larger for a $w_0w_a$ model.

Taken together, the DESI/CMB/SN datasets imply a 95\% credible range $\beta = 0.33 - 0.96$, providing weak evidence for scalar-field dark energy over $\Lambda$CDM.
The DESI and SN data alone favor a similar interval.
Because DESI+SN data only involve measurements at $z\simeq 0 - 2$,  evidence for scalar-field dark energy is independent of any deviations from $\Lambda$CDM at redshifts $z\gg 2$.

\subsection{Other work}

To put our work in context with contemporaneous studies \cite{2404.14341,2404.05722,2405.03933,2405.17396,2405.18747,2408.07175,2502.06929,2505.18937,2506.13047,2506.21542,2507.03090,2408.17318}, we highlight their findings, some of which have overlap with ours. Most of the studies that have considered scalar field models find them to be comparable to or slightly better than $\Lambda$CDM, though generally not as good as the best $w_0w_a$ models \cite{2404.14341,2405.17396,2405.18747,2502.06929,2505.18937,2506.13047,2506.21542,2507.03090}. 

Other studies find instead no significant evidence for evolving dark energy \cite{2405.03933}, and conclude that $\Lambda$CDM remains the preferred cosmological model \cite{2408.17318}. For example, Ref.~\cite{2408.07175} has argued that the step-like change in distances around $z\simeq 0.5$ associated with the peaked dark energy is an artifact of a discrepancy between the DESI and SN distance scales, not evidence for changing dark energy.

Ref.~\cite{2506.21542} focused on the implications of DESI for the fate of the Universe.  Ref.~\cite{2502.06929} proposed a different, less accurate one-parameter — the value of $w$ today — representation of different scalar potentials. The authors of Ref.~\cite{2404.05722} mapped constraints on $w_0w_a$ to $V$ and $V^\prime$, in the context of the Swampland conjecture. 
Several authors addressed alternatives to $w_0w_a$ that would avoid $w<-1$ \cite{2505.18937,2506.13047,2405.03933}, and one paper addressed the coupling of dark matter and dark energy in a string-inspired dark energy potential \cite{2507.03090}.

\subsection{Coda}

The hints of evolving dark energy are exciting, but $\Lambda$CDM is still a good fit to the DESI data, as well as to a host of other cosmological data.  
However, he DESI data also 
have a robust preference for a sharply-peaked dark energy, as described by a $w_0w_a$ model.   
This puzzling fact is either telling us something important about dark energy, a missing component of the energy density of the Universe, or about the DESI data themselves.
As more data are accumulated and as the DESI data are better understood, the 
``dark secrets'' that DESI is revealing should be clarified.

\vskip 20 pt
We thank Dragan Huterer for helping us understand the DESI dataset and for other useful suggestions, and Josh Frieman, Adam Riess, and Dan Scolnic for useful conversations.


\end{document}